\documentclass[epsf]{mn2e}

\usepackage{times}
\def\spose#1{\hbox to 0pt{#1\hss}}
\def\lta{\mathrel{\spose{\lower 3pt\hbox{$\mathchar"218$}}
     \raise 2.0pt\hbox{$\mathchar"13C$}}}
\def\gta{\mathrel{\spose{\lower 3pt\hbox{$\mathchar"218$}}
     \raise 2.0pt\hbox{$\mathchar"13E$}}}

\usepackage{graphicx}
\title[S\'ersic Properties of Remnants]{S\'ersic Properties
 of Disc Galaxy Mergers}
\author[Aceves, Vel\'azquez \& Cruz]{H. Aceves\thanks{E-mail:
aceves@astrosen.unam.mx}, H. Vel\'azquez, F. Cruz \\
Instituto de Astronom\'{\i}a, Universidad Nacional Aut\'onoma de 
M\'exico, Apdo. Postal 877, Ensenada,  BC 22800, M\'exico}

\begin{document}

\date{Accepted ------. Received ------; in 
original form ------}

\pagerange{\pageref{firstpage}--\pageref{lastpage}} \pubyear{2004}
\maketitle
\label{firstpage}

\begin{abstract}
S\'ersic parameters characterising the density profiles of remnants formed in 
collision-less disc galaxy mergers are obtained; no bulge is included in our 
simulations.  
For the luminous component we find that the 
  S\'ersic index is  $n\!\in\!(1.5,5.3)$ with 
$\langle n \rangle\approx 3 \pm 1$ and an effective radius of 
$R_{\rm e}\!\in\!(1.6,12.9)\,$kpc with 
$\langle R_{\rm e} \rangle \approx 5\pm 3\,$kpc. The mean values of these
quantities increases as the radial interval of fitting is reduced.
 A strong  correlation
 of $n$ with the central projected density $I_0$ is found 
($n\propto I_0^{-0.14}$) which is consistent with observations. 
No positive linear correlation between the size ($R_{\rm e}$) and structure
 ($n$) of our remnants is found; we do not advocate the existence of this.
The photometric plane (PHP)  of the luminous component 
 ($n\propto R_{\rm e}^{0.05}I_0^{0.15}$)
 agrees well, within the uncertainties and the assumption of a constant 
mass-to-light ratio, with those observationally determined for ellipticals
 in  the $B$-band ($n\propto  R_{\rm e}^{0.09}I_0^{0.15}$) and 
for $K$-band remnants ($n\propto  R_{\rm e}^{0.11}I_0^{0.14}$).
  We found that the surface defined by S\'ersic parameters 
$\{n,R_{\rm e},\mu_0\}$ in log-space is not a true plane, but a pseudo-plane
 with a small curvature at low values of $n$ owed to intrinsic properties of 
the S\'ersic model.
 The dark haloes of the  remnants have a 3-dimensional S\'ersic index of
$\langle n \rangle \approx 4 \pm 0.5$ that are smaller than the ones
 obtained for dark 
haloes in $\Lambda$CDM cosmologies $n \approx 6\pm 1$.
A tight dark S\'ersic ``plane'' (DSP) is also defined by the parameters of the remnants haloes with $n\propto r_{\rm e}^{0.07} \rho_0^{0.10}$.
  We conclude that collision-less merger remnants of {\it pure} disc galaxies
have S\'ersic properties and correlations consistent with those of observed 
in early-type galaxies and local remnants. It seems that a ``primordial'' bulge
 in spirals is not a necessary condition to form bona fide 
ellipticals on grounds of the  S\'ersic properties of remnants. 
\end{abstract}
%%%%%%%%%%%%%%%%%%%%%%%%%%%%%%%%%%%%%%%%%%%%%%%%%%

%%%%%%%%%%%%%

%\baselineskip=15pt

\section{Introduction}
%%%%%%%%%%%%%%%%%%%%%%%%%

Hierarchical galaxy formation theory (e.g. Cole~et~al.~2000, 
De~Lucia~et~al.~2005, Bower~et~al.~2005) considers that early-type galaxies
have an accretion/merger origin, as was originally suggested by Toomre (1977).
Observational  (e.g. Schweizer~1998, Struck~2005, Rothberg \& Joseph~2006, Kaviraj~et~al.~2006)  and theoretical (e.g.  Naab \& Burkert 2003,
Meza~et~al.~2005, Naab~et~al.~2005)  evidence supports this picture 
 although several topics remain 
unsolved (e.g. Peebles~2002, Tantalo \&
Chiosi~2004).

Early-type galaxies show several correlations among 
 their colours, luminosities,
velocity dispersions, effective radii and surface brightness  (e.g. 
Baum~1990, Faber~\&~Jackson~1976, Kormendy~1977, Djorgovski~\& Davis~1987,
Dressler~et~al.~1987, Bernardi~et~al.~2003). These
 correlations provide
constraints to any theory of  formation and evolution of
these galaxies.
Furthermore, their properties 
are linked with the 
 distribution of luminous and dark matter,  that would be 
 important when comparing with models of formation of
 elliptical galaxies.

Observational studies [e.g. Caon, Capaccioli \& D'Onofrio 1993 (CCD93),
 Graham \& Colles~1997, Binggeli \& Jerjen 1998, D'Onofrio~2001 (D01),
 Trujillo~et~al.~2004] have found that the surface 
brightness density profiles
 of early-type galaxies are better described by  a S\'ersic (1968)
 $R^{1/n}$--profile than  the classical de~Vaucouleurs (1948)
 $R^{1/4}$--profile. The index $n$ is  directly related with
 the cur\-va\-ture and
 ``concentration'' of the light profile (Trujillo, Graham \& Caon~2001).

Several observational relationships have been found between 
 the index $n$ and, for example, the 
 total luminosity ($L$), 
effective radius ($R_{\rm e}$)
 and central velocity dispersion (e.g. CCD93, Prugniel \& Simien~1997,
 Graham \& Guzm\'an~2003). 
Also, it has been found a  linear relation among
 $\log n$, $\log R_{\rm e}$ and
 $\mu_0$ (central brightness) termed the Photometric Plane (PHP) 
  for early-type galaxies [e.g. Khosroshahi~et~al.~2000 (K00), Graham~2002], 
 analogous to the Fundamental Plane (Djorgovski \& Davis 1987, 
 Dressler~et~al.~1987). Recently, 
Rothberg~\&~Joseph (2004, RJ04) have found  that nearby merger remnants 
have a peak in the $n$-distribution 
 at $n\approx 2$ with most values in the range of $1<n<6$, and in
some cases it is found that $n > 8$.

On other hand, theoretical studies of  S\'ersic properties of merger
remnants have appeared recently.  For example, 
G\'onzalez-Garc\'{\i}a~\&~Balcells (2005, GGB)
and  Naab~\&~Trujillo~(2005, NT) find in 
 collision-less simulations  that 
bulge-less progenitors lead to  ranges of
$n\! \in\! (2.4,3.2)$ and $(1.2,3.1)$, respectively; when 
 a single S\'ersic function is used to fit the entire remnant. 
  For progenitors with a bulge component they  obtain 
about the same range of S\'ersic index, 
 $n\! \in \!(3,8)$. 
 Since ``bona fide'' ellipticals have values $n\gta 4$, they
reach the conclusion
that collision-less merger remnants of pure disc galaxies do not
lead to concentrations, indicated by $n$, similar to those
found in intermediate or giant elliptical galaxies (e.g. Graham~et~al.~1996).

The above findings suggest that a 
  primordial bulge in spirals is a necessary condition to 
form bona fide ellipticals in the hierarchical merging scenario.
 However, we show below,  collision-less mergers of pure discs
 can cover the range of observed values of the shape parameter $n$,
 and can reproduce adequately other observational correlations.

  S\'ersic model in a de-projected form has been recently used 
to represent the dark matter distribution in $\Lambda$CDM haloes
(Navarro~et~al.~2004, 
Merritt~et~al.~2005, Graham~et~al.~2005, Prada~et~al.~2005), in order
to have a better estimation of the inner asymptotic logarithmic derivative. 
 A mean value of a 3D S\'ersic
index $\approx 6$, with a scatter of $\approx 1$, has been found 
in these works. 
So it is of interest to determine the three-dimensional S\'ersic 
 parameters that characterise  our remnants.

In this work, we study the structural properties of remnants
as provided by fitting  a  S\'ersic profile to their luminous and dark 
mass distribution.
The paper has been organised as follows: in $\S$\ref{sec:model} we present a
summary of the properties of our progenitors, some details
of the simulations performed, as well as some basic characteristics
 of  S\'ersic profile; 
both projected and deprojected. In $\S$\ref{sec:results}
 we present  distributions and
correlations, in two and three-dimensions, found among the  different 
S\'ersic parameters for our remnants, and compare them with observations.
 S\'ersic properties of the dark haloes of the remnants are 
determined, some correlations presented,  and 
compared with those obtained in cosmological simulations.
 Some final comments are given in $\S$\ref{sec:discussion} 
and a summary of our conclusions.

%%%%%%%%%%%%%%%%%%%%%%%%%%%%%%%%%%%%%%%%%%%%
\section{Simulations and S\'ersic Functions}\label{sec:model}
%%%%%%%%%%%%%%%%%%%%%%%%%%%%%%%%%%%%%%%%%%%%

\subsection{Galaxy models}
%%%%%%%%%%%

The galaxy models used in this work have been already described in Aceves \&
Vel\'azquez (2005) and follow the method outlined by 
 Shen, Mo \& Shu (2002) to
obtain the global properties of the discs, once the haloes properties are
known. Our numerical galaxies do \emph{not} include a bulge-like component. 
 The dark haloes
follow a modified NFW (Navarro, Frenk \& White~1997) model with an exponential
cutoff. The discs have a typical  exponential density profile, and satisfy
the Tully-Fisher relation at redshift $z\!=\! 1$; roughly a look-back time of 
$8\,$Gyr in a $\Lambda$CDM cosmology with Hubble parameter $h=0.7$. 
Only discs satisfying the Efstathiou, Lake \& Negroponte~(1982) stability 
criterion were used.

In this work, an additional simulation to those reported in Table~1 of
 Aceves \&~Vel\'azquez
(2005) has been done. This is a merger from the resulting remnants of
 $M01$ and $M05$, label  $MM$. 
All simulations were carried out using a parallel version of {\sc
  gadget}-1.1 code, a tree base code (Springel, Yoshida \& White 2001), and 
evolved  for $\approx \! 8\,$Gyr with conservation of energy better than
$0.25$ percent.

\subsection{Density Profiles}
%%%%%%%%%%%%%%%%%%%%%%%%%%%%%%%%%%%%%%

We  fit only S\'ersic profiles to our merger remnants; no bulge-disc
decomposition is attempted since progenitors lack any bulge component.
The S\'ersic surface luminous-mass density profile is given by
\begin{equation}
\Sigma(R) = \Sigma_0 \,  {\rm e}^{ -b (R/R_{\rm e})^{1/n}}\;,
\label{eq:sersic}
\end{equation}
where $R$ is the projected spherical radius, $R_{\rm e}$ is the effective
 radius,  $n$ the index of the profile, $b=b(n)\approx 2n - 0.324$ (Ciotti
 \& Bertin 1999) and $\Sigma_0$ the central surface density. 
Index $n$ is associated with the curvature and the concentration of the 
profile  (Trujillo, Graham \& Caon~2001); 
 $n\!=\!1$ corresponds to an exponential
 profile while the classical de Vaucouleurs (1948)
 profile is obtained for $n=4$. 
 
 The accumulated projected luminous mass, $M_{\rm
  L}(R)$, is given by
\begin{equation}
M_{\rm L}(R)= \int_0^R \Sigma(R) {\rm d}(\pi R^2) = \frac{2\pi n  \gamma(\alpha,x)}{b^{2n}}\,  \Sigma_0 R^2_{\rm e} \,,
\end{equation}
where $\alpha \equiv 2n$, $x \equiv b(R/R_{\rm e})^{1/n}$, and
$\gamma (\alpha,x)$ is the incomplete gamma function.
  The total projected luminosity mass is given by
\begin{equation}
M_{\rm L} =  \frac{2\pi n}{b^{2n}} \Gamma(2n) \, \Sigma_{\rm 0} R^2_{\rm e}\, ,
\label{eq:Lmass}
\end{equation}
being $\Gamma(\alpha)$ the complete gamma function. 
A summary of S\'ersic projected profile properties 
 is given by Graham \& Driver (2005). When comparing our simulations with
 observations we assume a constant mass-to-light ratio, so that $\Sigma
 \!\propto \! I$; where $I$ refers to the surface brightness.

 The three-dimensional (3D) S\'ersic profile  
  is 
\begin{equation}
\rho(r) = \rho_{\rm 0} \,  {\rm e}^{ -d (r/r_{\rm e})^{1/n}}\;,
\label{eq:3Dsersic}
\end{equation}
where $r$ is the spatial radius, $d\approx  3n-1/3+0.005/n^2$
 (Graham~et~al.~2005) such that
$r_{\rm e}$ is the half-mass spatial radius. The total mass is determined 
from 
\begin{equation}
M_{\rm t} =  \frac{4\pi n}{d^{3n}} \Gamma(3n) \, \rho_0 r^3_{\rm e} \,.
\label{eq:3Dmass}
\end{equation}

S\'ersic parameters for 
the luminous component were computed along 400 
different random line-of-sights. To each projection a circularly averaged
density profile $\Sigma(R)$ was determined, and a S\'ersic profile
(\ref{eq:sersic}) fitted by $\chi^2$--minimisation using the
Levenberg-Marquardt method (Press~et~al.~1992) to obtain  $\{n,R_{\rm
  e},\Sigma_0\}$. S\'ersic parameters for dark haloes are obtained by a similar
procedure, but using equation (\ref{eq:3Dsersic}).

\subsubsection{Fitting Range}
%%%%%%%%%%%%%%%%%%%%%%%%%%%%%%%%%%%%

The fitting  set of parameters  depend on the
methodology used to obtain them. In particular, there have been indications
that  these  parameters depend  on both 
the covered range of surface brightness range 
(e.g. Capaccioli, Caon \& D'Onofrio 1992) and the spatial 
radial interval for fitting  (e.g. Kelson~et~al.~2000).

Also, the determination of fitted parameters degrades when
the inner parts of a galaxy are not well considered. For example, the index $n$
tends more to be a representation of the outer slope of the profile than of
the curvature of the luminosity distribution (Graham~et~al.~1996).
The treatment and quality of data has also an effect on the fitted parameters.
 For example, CCD93 obtain higher values of $n$ for NGC~4406, NGC~4552 and 
NGC~1399 (14.9, 13.9, 16.8) in comparison with D01 ($6.5$, $7.2$, $6.1$).

We have considered two radial intervals for our fits in
 order to asses their effect on the S\'ersic parameters.
The first radial interval, $I_1$, is taken from our numerical resolution value
 $\xi_{\rm i}=100\,$pc to the outer radius $\eta_{95}$, which encloses 95 percent of the projected luminous mass and is 
determined directly from the simulations; thus,  
 $I_1=[\xi_{\rm i},\eta_{95}]$. 
 The second one, $I_2$, uses another inner point at 
$\xi_{\rm f}=10\xi_{\rm i}$,\footnote{ \footnotesize For reference, in a $\Lambda$CDM cosmology with $h=0.7$ we have 
that $1''=464\,{\rm pc}$ at the distance of the Coma cluster
 ($z\!=\!0.023$), $977\,$pc at $z\!=\!0.05$, and $4.5\,$kpc at $z\!=\!0.3\,$.}
 and outer point at $\eta_{70}$; this enclosing 70 percent of the luminous mass. For each line-of-sight used,  two uniform random numbers $\xi \in [\xi_{\rm i},\xi_{\rm f}]$ and 
$\eta \in [\eta_{70},\eta_{95}]$ are generated that 
in turn define $I_2=[\xi,\eta]$. In the Appendix we discuss some effects  the radial range of a fit has on the parameters estimated using synthetic models.

%%%%%%%%%%%%%%%%%%%%%%%%%%%%%%%%%%%%%%%%%%%%%%%%%%%%%%%%%%%%%%%%%%%%%%%%%
\section{Results}\label{sec:results}
%%%%%%%%%%%%%%%%%%%%%%%%%%%%%%%%%%%%%%%%%%%%%%%%%%%%%%%%%%%%%%%%%%%%%%%%

In this section we present the results of the fittings done,
 both ``luminous'' and dark, to the merger remnants, as well
as several relationships among them based in observational studies. 

Table~\ref{tab:global}  lists different
 global physical properties of our
remnants obtained directly from the $N$-body simulations. 
Column (2) is the total half-mass radius $R_{\rm h}$,
(3) the virial radius $R_{\rm v}$, (4) the virial velocity $V_{\rm v}$,  
(5) the total luminous mass ${\rm M}_{\rm lum}$ and (6) 
 the total bounded mass
${\rm M}_{\rm tot}$, and column (7) is the virial ratio at the end of
the simulation. The last column (8) provides the ratio of the total mass
 of the secondary to the primary galaxy in the simulations.
The merger labelled as $MM$ corresponds to  the simulation where the 
resulting remnants of $M01$ and $M05$ were merged together in a parabolic encounter.

%%%%%%%%%%%%%%%%%%%%%%%%%
\begin{table}
% \centering
 \begin{minipage}{80mm}%{140mm}
  \caption{Physical properties of remnants}\label{tab:global}
  \begin{tabular}{lcrrrrll}
  \hline
{\sc id}& $R_{\rm h}$ & $R_{\rm v}$ & $V_{\rm v}$ & 
$ \frac{{\rm M}_{\rm lum}}{10^{10}}$ 
& $ \frac{ {\rm M}_{\rm tot}}{10^{11} }$ &  $\frac{2T}{|W|}$ & $\frac{2}{1}$\\
        &  [kpc]   & [kpc] & [km/s] & [M$_\odot$] &  [M$_\odot$] &  & \\
 \hline
 $M01$  & 66.9 & 156.1 & 213.0 & 10.00 & 16.60 & 0.99 & 0.32\\
 $M02$  & 29.8 &  71.4 & 108.5 &  0.60 & 1.95 & 0.99 & 0.46 \\
 $M03$  & 24.6 &  56.4 &  99.0 &  0.54 & 1.29 & 0.99 & 0.53\\
 $M04$  & 41.6 &  96.2 & 132.8 &  1.55 & 3.98 & 0.99 & 0.74\\
 $M05$  & 22.2 &  48.6 & 100.3 &  0.83 & 1.15 & 1.00 & 0.93\\
 $M06$  & 27.3 &  63.4 &  96.9 &  0.81 & 1.41 & 0.99 & 0.87\\
 $M07$  & 24.0 &  55.3 & 105.8 &  1.02 & 1.45 & 0.99 & 0.51\\
 $M08$  & 33.8 &  80.5 &  92.5 &  0.37 & 1.62 & 0.98 & 0.97\\
 $M09$  & 28.7 &  66.1 & 103.5 &  1.41 & 1.66 & 0.99 & 0.98\\
 $M10$  & 33.3 &  74.9 & 110.8 &  1.66 & 2.19 & 0.99 & 0.70\\
 $M11$  & 32.4 &  76.6 & 178.2 &  4.47 & 5.62 & 1.00 & 0.14\\
 $M12$  & 32.1 &  74.9 & 147.0 &  2.39 & 3.72 & 1.01 &0.18\\
 $MM$  & 68.2 & 163.1 & 216.7 & 10.72 & 17.78 & 1.02 & 0.07 \\
\hline
\end{tabular}
\end{minipage}
\end{table}
%%%%%%%%%%%%%%%%%%%%%%%%%%%%%%%5

%%%%%%%%%%%%%%%%%%%%%%%%%5
\begin{table}
 \centering
 \begin{minipage}{140mm}
  \caption{Mean parameters using radial range $I_1$}\label{tab:fits1}
  \begin{tabular}{lcrccc}
  \hline
{\sc id}& $n$ & $R_{\rm e}$ & $-\mu_0$ & $-M_{T}$ & {\sc rms} \\
        &     & [kpc] & [${\rm M}_\odot/{\rm kpc}^2$] & [M$_\odot$] & \\
 \hline
$M01$ & $ 4.3 \pm   0.4$ & $ 9.2 \pm   1.9$ & $28.5 \pm    0.5$ & $27.6$ & $0.10$ \\ 
$M02$ & $ 2.1 \pm   0.1$ & $ 2.7 \pm   0.2$ & $23.3 \pm    0.3$ & $24.4$ & $0.10$ \\ 
$M03$ & $ 1.9 \pm   0.1$ & $ 2.5 \pm   0.0$ & $23.0 \pm    0.1$ & $24.3$ & $0.12$ \\ 
$M04$ & $ 2.8 \pm   0.1$ & $ 3.9 \pm   0.2$ & $25.1 \pm    0.2$ & $25.5$ & $0.14$ \\ 
$M05$ & $ 3.9 \pm   0.2$ & $ 1.7 \pm   0.1$ & $28.4 \pm    0.4$ & $24.8$ & $0.10$ \\ 
$M06$ & $ 2.5 \pm   0.1$ & $ 4.2 \pm   0.3$ & $23.6 \pm    0.4$ & $24.7$ & $0.12$ \\ 
$M07$ & $ 3.1 \pm   0.2$ & $ 2.6 \pm   0.3$ & $26.1 \pm    0.5$ & $25.0$ & $0.10$ \\ 
$M08$ & $ 2.6 \pm   0.1$ & $ 2.1 \pm   0.2$ & $24.5 \pm    0.3$ & $23.9$ & $0.12$ \\ 
$M09$ & $ 2.7 \pm   0.2$ & $ 8.1 \pm   0.8$ & $23.2 \pm    0.6$ & $25.3$ & $0.20$ \\ 
$M10$ & $ 3.2 \pm   0.2$ & $ 6.4 \pm   0.7$ & $24.8 \pm    0.6$ & $25.5$ & $0.14$ \\
$M11$ & $ 1.6 \pm   0.1$ & $ 9.3 \pm   1.4$ & $22.0 \pm    0.5$ & $26.6$ & $0.08$ \\ 
$M12$ & $ 3.2 \pm   0.2$ & $ 4.3 \pm   0.6$ & $26.0 \pm    0.6$ & $25.9$ & $0.19$ \\ 
$MM$ & $ 2.4 \pm   0.1$ & $ 9.1 \pm   0.7$ & $24.5 \pm    0.1$ & $27.6$ & $0.19$ \\ 
\hline
\end{tabular}
\end{minipage}
\end{table}
%%%%%%%%%%%%%%%%%%%%%%%%%

%%%%%%%%%%%%%%%%%%%%%%%%%
\begin{table}
 \centering
 \begin{minipage}{140mm}
  \caption{Mean parameters using random radial range $I_2$}\label{tab:fits2}
  \begin{tabular}{lcrccc}
  \hline
{\sc id}& $n$ & $R_{\rm e}$ & $-\mu_0$ & $-M_{T}$ & {\sc rms} \\
        &     & [kpc] & [${\rm M}_\odot/{\rm kpc}^2$] & [M$_\odot$] & \\
 \hline
$M01$ & $ 5.8 \pm   1.3$ & $13.2 \pm   5.5$ & $30.9 \pm    2.0$ & $27.8$ & $0.05$ \\
$M02$ & $ 2.1 \pm   0.3$ & $ 2.4 \pm   0.3$ & $23.7 \pm    0.8$ & $24.4$ & $0.05$ \\ 
$M03$ & $ 2.2 \pm   0.4$ & $ 2.3 \pm   0.2$ & $23.9 \pm    0.8$ & $24.3$ & $0.05$ \\ 
$M04$ & $ 3.3 \pm   1.0$ & $ 3.1 \pm   0.5$ & $26.5 \pm    2.1$ & $25.4$ & $0.07$ \\ 
$M05$ & $ 3.2 \pm   1.1$ & $ 2.1 \pm   0.6$ & $26.6 \pm    2.1$ & $24.8$ & $0.04$ \\ 
$M06$ & $ 2.7 \pm   0.4$ & $ 3.7 \pm   0.3$ & $24.3 \pm    0.9$ & $24.7$ & $0.06$ \\ 
$M07$ & $ 2.6 \pm   0.7$ & $ 3.0 \pm   1.0$ & $24.8 \pm    1.4$ & $25.0$ & $0.06$ \\ 
$M08$ & $ 2.1 \pm   0.3$ & $ 2.3 \pm   0.5$ & $23.3 \pm    0.8$ & $23.9$ & $0.04$ \\ 
$M09$ & $ 2.9 \pm   0.5$ & $ 6.9 \pm   0.9$ & $23.8 \pm    1.2$ & $25.3$ & $0.10$ \\ 
$M10$ & $ 3.5 \pm   0.6$ & $ 5.8 \pm   0.7$ & $25.7 \pm    1.4$ & $25.5$ & $0.06$ \\ 
$M11$ & $ 1.6 \pm   0.3$ & $ 9.6 \pm   1.9$ & $22.0 \pm    0.8$ & $26.6$ & $0.06$ \\ 
$M12$ & $ 3.2 \pm   1.4$ & $ 6.3 \pm   4.0$ & $25.5 \pm    2.4$ & $26.0$ & $0.16$ \\ 
$MM$ & $ 3.2 \pm   0.6$ & $ 9.2 \pm   1.5$ & $26.3 \pm    1.0$ & $27.6$ & $0.07$ \\
\hline
\end{tabular}
\end{minipage}
\end{table}
%%%%%%%%%%%%%%%%%%%%%%%%%%%%%%%5

Tables~\ref{tab:fits1} and~\ref{tab:fits2} summarise the mean values of 
 the fitted S\'ersic parameters $\{n,R_{\rm e},\mu_0\}$ 
($\mu_0\!=\!-2.5\log \Sigma_0$), 
the total ``magnitude''
 ($M_T\!\equiv\! -2.5\log M_{\rm L}$) and the  {\sc rms} of 
the fit, for the different projections for both radial intervals $I_1$ 
and $I_2$; respectively.
 Here $M_{\rm L}$ is
 determined from the fitted values using equation (\ref{eq:Lmass}). 
Standard deviations are listed for the S\'ersic parameters.
The values of $M_{\rm L}$ determined from the
fits agree very well with $M_{\rm lum}$.

\subsection{Luminous Distributions}
%%%%%%%%%%%%%%%%%%%%%%%%%%

\subsubsection{Shape parameter}\label{ssec:n}
%%%%%%%%%%%%%%%%%%%%%%%%%%%%%%%%

\begin{figure}
\centering 
\includegraphics[width=8cm,height=7cm]{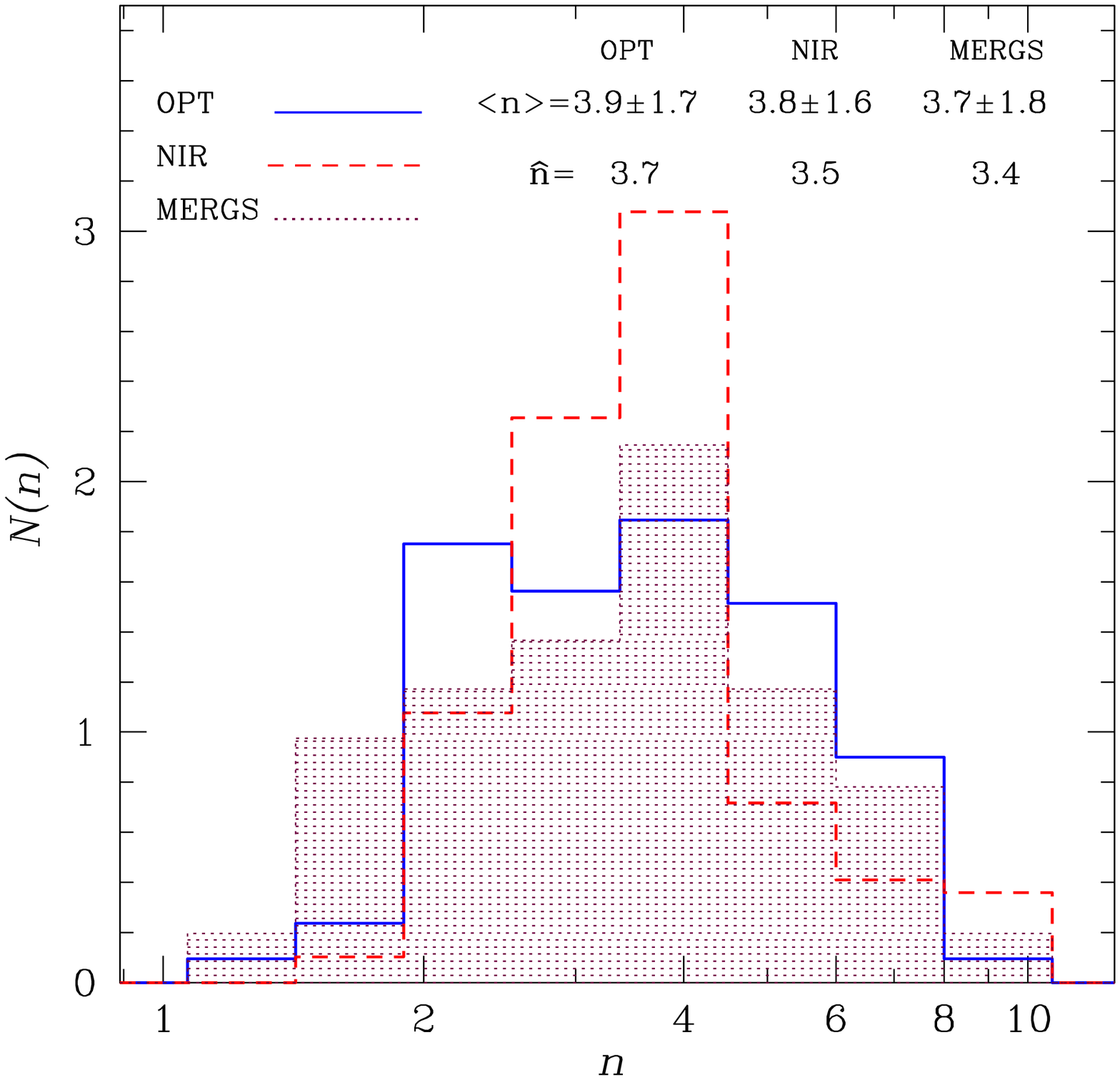}
\includegraphics[width=8cm,height=7cm]{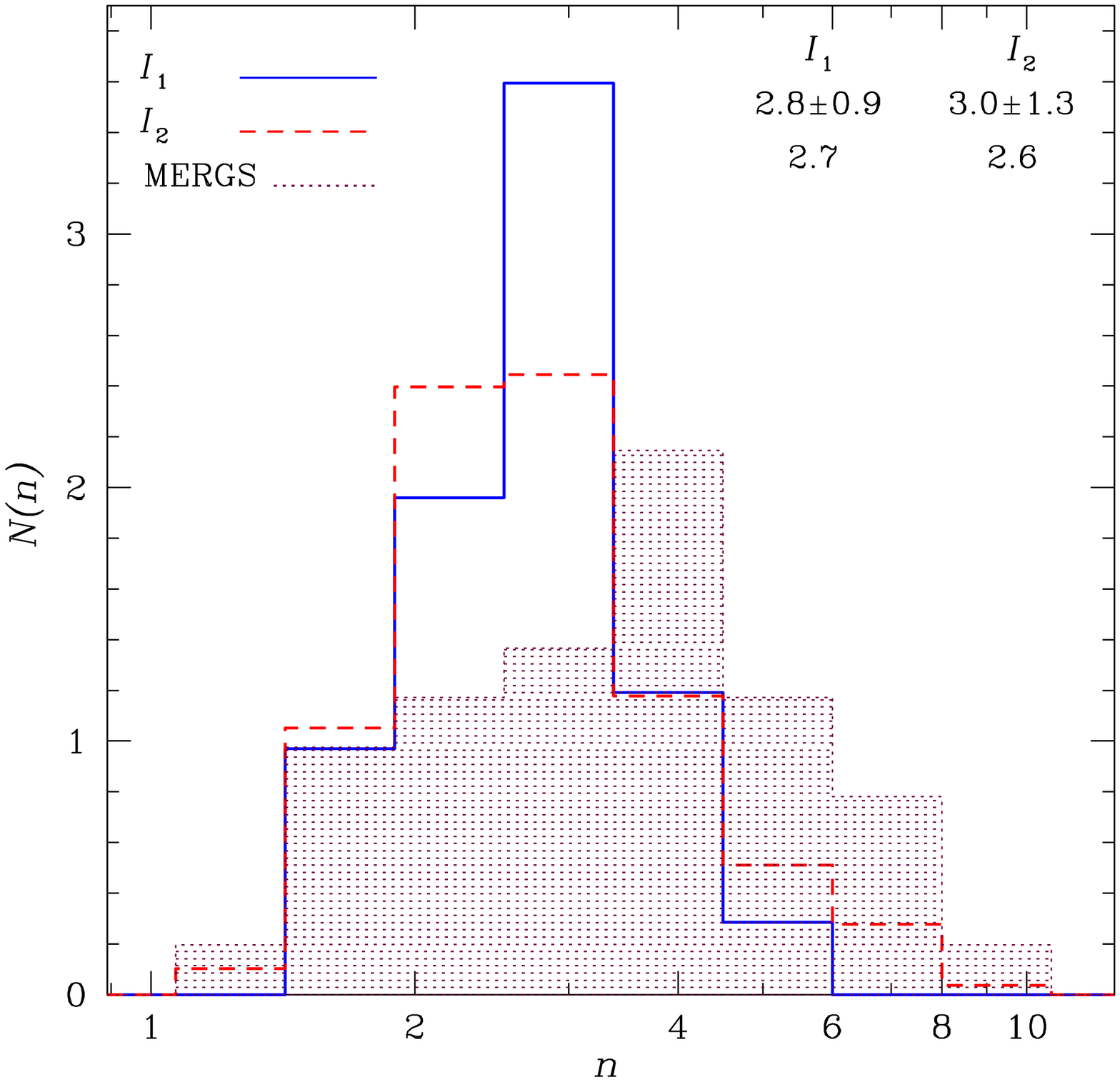}
\vspace{-0.3cm}
\caption{ ({\it Top}) 
Frequency distributions $N(n)$ for S\'ersic index $n$ of a 
sample of early-type galaxies in the optical ({\sc opt}, {\it solid line}) 
and near-infrared bands ({\sc nir}, {\it dashed line}).  
({\it Bottom}) Frequency distribution of our $N$-body remnants using both 
fitting intervals $I_1$ and $I_2$.
The mean values and their standard deviations are indicated; below these,
the median is  indicated. 
 $K$-band merger remnants ({\sc mergs}) of 
Rothberg \& Joseph (2004) correspond to the shaded histogram 
in both panels.
}
\label{fig:Nglxs}
\end{figure}

Figure~\ref{fig:Nglxs}~({\it top})
 shows the frequency distribution of $n$
 for a set of observational data in optical wave bands (D01,
 La~Barbera~et~al.~2005) and in the near-infrared ($K$) band
 [La~Barbera~et~al~2005, Ravikumar~et~al.~2005 (R05)]. 
 A total of 169 galaxies in the optical and 156 in the $K$
band were used here.
The frequency distribution of 41 merger remnants observed in the
$K$-band by 
Rothberg \& Joseph (2004) are also indicated as a shaded histogram.
The mean and standard deviations of these data sets are indicated, as 
well as their median.

%%%%
%%%

In  Figure~\ref{fig:Nglxs}~({\it bottom}) 
we show the distribution of $n$ 
for our merger remnants using the radial fitting intervals $I_1$ and $I_2$. 
The frequency distribution for our  $N$-body remnants  peak at a value
$n\approx 3$ in both cases; although  
using  $I_2$ it shows a somewhat broader distribution. 
For $I_1$ it is found that $n\in (1.5,5.3)$ and for $I_2$ that
  $n\! \in \! (1.4,9.5)$. These values are in good agreement with those found
  in intermediate mass ellipticals (e.g. Graham~\&~Guzm\'an~2003, de
  Jong~et~al.~2004, Trujillo, Burkert \& Bell~2004, Ellis~et~al.~2005), 
  some brightest cluster galaxies (e.g. Graham~et~al.~1996),  dwarf ellipticals
  (e.g. Binggeli \& Jerjen~1998, Young \& Currie 2001), and the local merger remnants of RJ04.

Our results using $I_1$, and \emph{no} bulge, are consistent with the values found by
 NT and GGB for their models with a bulge in the progenitors. Furthermore, 
using interval $I_2$ lead to some values $n \! \approx \! 9$. 
This does not seem to be due to the methodology in the
 computation of the surface density profiles. 
NT construct artificial
images analogous to the observational procedure while GGB fit ellipses  
 to isodensity contours,
 both considering a wide range in the radial fitting range, 
 and obtaining similar ranges for $n$. 
It is likely that 
 differences in the way models of the progenitors are set up be  
 probably one of the 
 reasons behind the differences with our results; see
 $\S$\ref{sec:discussion}.

%%%%
\begin{figure}
\centering
\includegraphics[width=8cm,height=7cm]{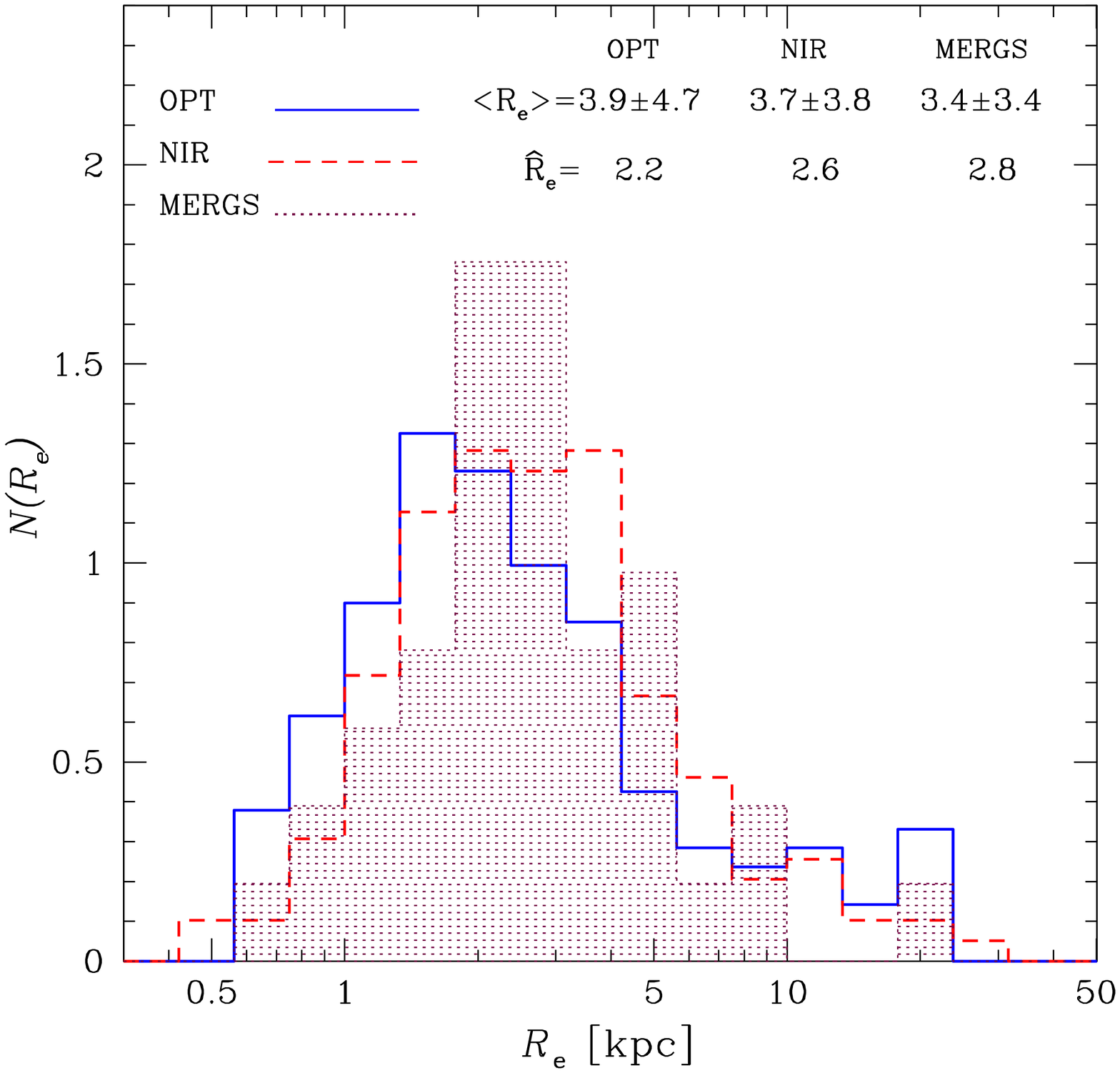}
\includegraphics[width=8cm,height=7cm]{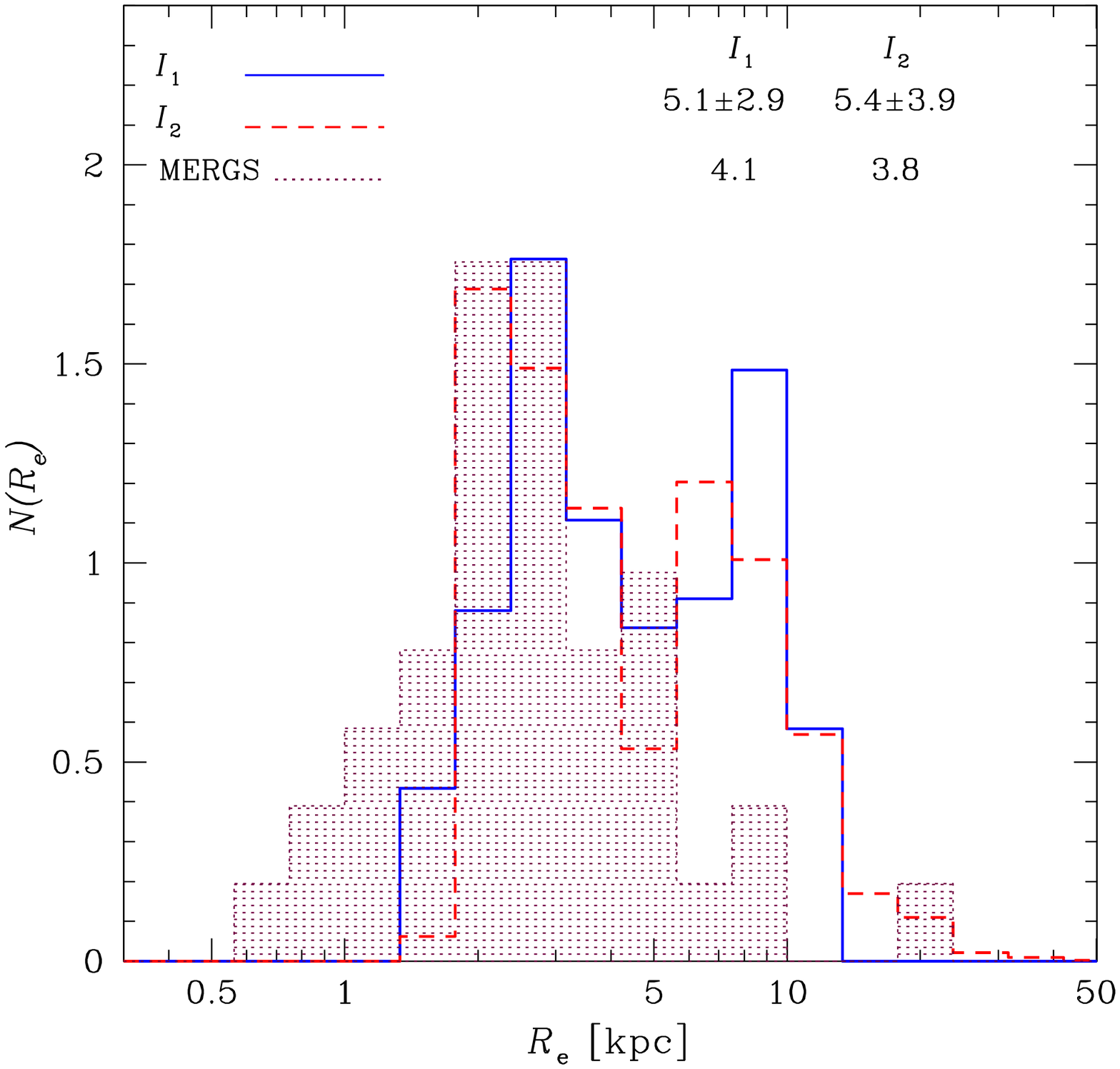}
\vspace{-0.3cm}
\caption{Frequency distributions of $R_{\rm e}$ for observed early-type 
galaxies ({\it top}) and $N$-body
remnants ({\it bottom}). Mean values and dispersion, along with medians, are
indicated as in Figure~\ref{fig:Nglxs}. The same observational 
data sets considered in Figure~\ref{fig:Nglxs} are used here.
}
\label{fig:Nre}
\end{figure}
%%%%

\subsubsection{Effective Radius}
%%%%%%%%%%%%%%%%%%%%%%%%%%%%%%%%

Figure~\ref{fig:Nre} ({\it top}) shows  the
observed frequency distribution of effective radii $R_{\rm e}$ for the 
data considered in $\S$\ref{ssec:n} and 
 that corresponding to our remnants ({\it bottom}). 
For the fitting radial range $I_1$ we obtain 
 $R_{\rm e}\! \in \!(1.6,12.9)\,$kpc, and for $I_2$ we have
$R_{\rm e}\! \in \! (1.6,34.5)\,$kpc. 
The average value of the observational data  is about
 $4\,$kpc  and for our remnants is about
$5\,$kpc.  It can be noticed that $R_{\rm e}$ shows a larger dispersion of
values than the index $n$ depending on the fitting interval. This was also noticed by Binggeli \& Jerjen (1998).

Our remnants have a lower bound of
 $R_{\rm e}\!\approx \!1.5\,$kpc, while 
the observational data considered here can reach smaller values
 $R_{\rm e} \! \gta \! 0.5\,$kpc.  We
are not able to reproduce the small values of $R_{\rm e}$ mainly because 
 in our sample of initial conditions  no  pairs of small progenitors were
 included. 
  On other hand, values of $R_{\rm e}\gta 10\,$kpc can be reproduced by
our more massive remnants ($M01$ and $MM$); see 
Tables~\ref{tab:fits1} and~\ref{tab:fits2}.

A unique comparison with the distribution of$R_{\rm e}$  values  found by
NT and GGB is not possible, since their models can be scaled  to arbitrary
physical units; a thing that is not possible here due to the way our disc 
galaxy  progenitors were built up. Nonetheless, if we use the range of
dimensionless values found by NT ($1\! < \! R_{\rm e} \! < \! 1.7$) 
for systems classified as pure ``bulges'', 
and use a length unit of $3.5\,$kpc  
 (i.e., the radial scale-length of the Milky Way)
 to transform their results to physical units,
we find that both results are consistent. 
Also, we obtain qualitatively the same behaviour
as the one shown in their Figure~18 where a sharp cut 
 at the lower-end of the distribution, as well as  an
extended tail at larger values.

 Considering the observational  values of $R_{\rm e}$ 
and those in  our $N$-body remnants, we can establish with confidence that the
simulations can reproduce quite well the observed range of values. Even some
large values of $R_{\rm e}$ found in giant ellipticals (e.g., 
Graham~et~al.~1996) are reproduced.

\subsection{Luminous Correlations}
%%%%%%%%%%%%%%%%%%%%%%%%%%
%%%%%%%%%%%%%%%%%%%%%%%%%%%

Several works (e.g. CCD93, D01, R05)
 have found a series of correlations among  S\'ersic 
parameters in early-type galaxies. We now turn to study some of these
 and compare them
with the properties of our numerical remnants. Firstly, we consider
 two-dimensional correlations, and then turn to consider the so called 
Photometric Plane (PHP) [e.g. K00].

%%%%
\begin{figure}
\centering
\includegraphics[width=8.5cm]{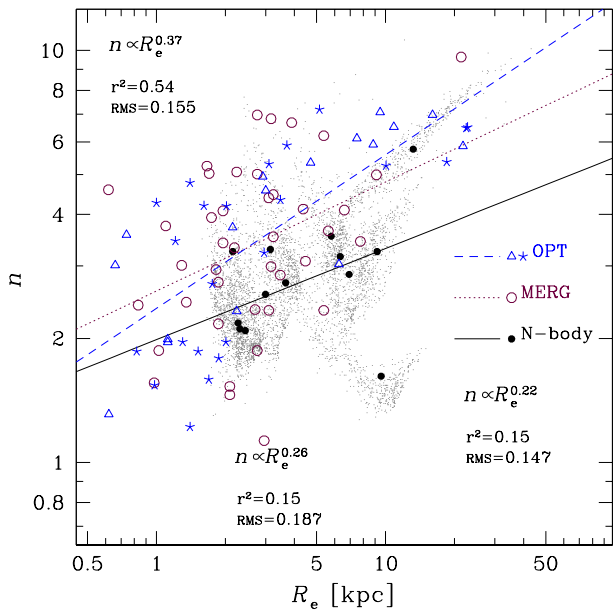}
\vspace{-0.5cm}
\caption{S\'ersic index $n$ versus effective radius for the observational data
of D01, R05, and RJ04. Genuine E galaxies from D01 are represented by
a triangle ($\triangle$), while those  classified as of an uncertain type 
by a star ($\star$). Merger remnants of RJ04 are open circles ($\circ$).
Lines correspond to least-square fits. The values obtained, 
along with the coefficient of determination and {\sc rms} in $\log n$ of 
the fit  are 
indicated.
The average results for our $N$-body remnants, using the radial range $I_2$,
 are shown with solid  dots ($\bullet$) and   
points correspond to different projections of those numerical remnants.
}
\label{fig:NRcorrel}
\end{figure}
%%%%

\subsubsection{Two Dimensional Correlations}\label{sec:2Dcorrel}
%%%%%%%%%%%%%%%%%%%%%%%%%%%%%%%

In the work of Caon~et~al.~(1993) it was stated that a linear positive
 correlation between $n$ and $R_{\rm e}$ exists for early-type galaxies; 
 they find that
$n\propto R_{\rm e}^{0.52}$ for early-type galaxies in Virgo. 
A similar conclusion was reached by
D'Onofrio, Capaccioli \& Caon (1994) analysing galaxies in Fornax.
Combining the data  of both works 
one finds $n\propto R_{\rm e}^{0.50}$ with
a Pearson's linear correlation 
coefficient $r=0.72$. 
The statement of CCD93 that 
structure (as indicated by $n$) of an elliptical 
depends on its size 
$R_{\rm e}$ has been supported 
by the analysis of Trujillo, Graham \& Caon (2001).

Figure~\ref{fig:NRcorrel} shows
index $n$ against $R_{\rm e}$ for some of 
the data considered here, as
well as the values obtained for our disc galaxy merger remnants. A
linear least-square fit to the data of D01 leads to   
$n\propto R_{\rm e}^{0.37}$, and for RJ04 mergers   
 $n\propto R_{\rm e}^{0.26}$;  with linear correlation 
coefficients $r=0.73$ and $0.39$, respectively. 
A similar fit
to our remnants yields $n\propto R_{\rm e}^{0.22}$ with $r=0.39$.

However, the observational 
data plotted in Figure~\ref{fig:NRcorrel} shows a large scatter
  around the assumed linear correlation; a fact already noticed by
other authors (e.g. Trujillo~et~al.~2001). These fluctuations 
are quantified by considering the {\sl coefficient of
determination} ($r^2$) that measures 
the proportion of the variance of one variable that is predictable
from the other (e.g. Ryan~1997). 
As indicated in Figure~\ref{fig:NRcorrel}, 
the coefficients of determination are rather small, and the {\sc rms} of the
fits are large, so its is not clear
that a {\it true} linear correlation exists
 between $\log n$ and $\log R_{\rm e}$.

We notice also 
that the $N$-body remnant $M05$, being the smallest one, has the second 
largest $n$ value in our simulations.  These results lead us to state that
there is {\it no} linear positive correlation between 
the ``structure'' and size of an elliptical. It seems that the values of $n$
  and $R_{\rm e}$ are restricted by some physical mechanism to a finite region
  of the S\'ersic parameter space; an option also indicated by 
Trujillo~et~al.~(2001).

%%%%%%
\begin{figure}
\centering
\includegraphics[width=8.5cm]{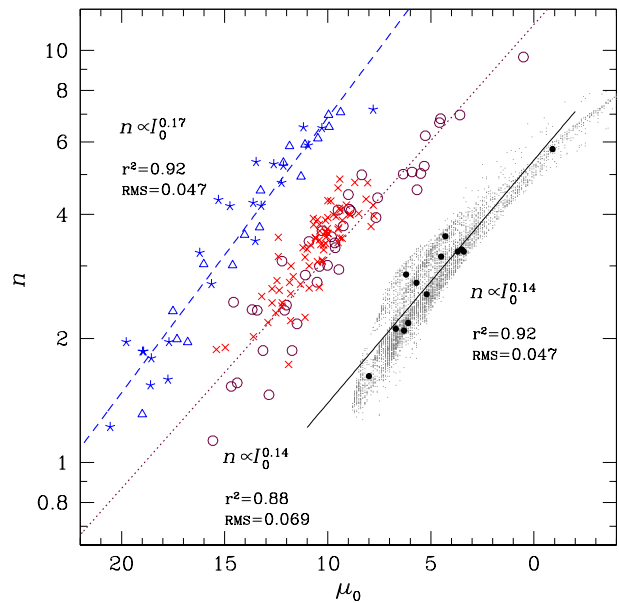}
\vspace{-0.5cm}
\caption{S\'ersic index versus $\mu_0$ from observational data of 
 D01 ($\triangle, \star$),  RJ04 ($\circ$), R05 ($\times$).  
and $N$-body remnants mean values ($\bullet$) and their projected ones
 ({\it points}).
Lines are the resulting scaling relations obtained from least-square fits.
The coefficient of determination ($r^2$) and the {\sc rms} are also indicated. 
For clarity,  results from $N$-body remnants have been
 displaced by a constant value along the $\mu_0$--axis.}
\label{fig:NMu0correl}
\end{figure}
%%%%

On other hand, a stronger observational correlation  
 has been found between $n$ and the central
brightness $\mu_0$ in ellipticals (e.g. K00, Graham \& Guzm\'an~2003) and 
it appears to extend to  dwarf ellipticals (e.g. Binggeli \& Jerjen 1998,
 R05). 
In Figure~\ref{fig:NMu0correl} we plot these quantities for the observational 
data of D01, R05, RJ04,  and for comparison our $N$-body remnants.
 A linear fit  $\log n$--$\mu_0$ to
these data leads to  $n\propto I_0^{0.17}$ for D01,  $n\propto I_0^{0.14}$
for both RJ04 and the ellipticals in R05, and $n\propto I_0^{0.14}$ 
 for our merger remnants. All fits have $r^2\gta 0.9$ and an
{\sc rms}~$\lta 0.07$, that lead us to conclude that $\log n$--$\mu_0$ is
 a true linear correlation, at least for the range of $n$ values considered.

As shown, the numerical remnants presented here 
are able to reproduce very well the $\log n$--$\mu_0$
 correlation and its tightness. We recall that we have assumed 
 a constant mass-to-light ratio to convert  $\Sigma_0$ to $I_0$ in
order to compare with  observations. 
Hence, it appears that at least in the range
of masses of our remnants (Table~\ref{tab:global}), there is no need 
to assume a dependence of the mass-to-light ratio  dependence with
 mass (or luminosity) to reproduce the observations.

\subsubsection{The Photometric ``Plane''}
%%%%%%%%%%%%%%%%%%%%%%%%%%%%%%%%%%%%%%%%%

Several authors [e.g. K00, Graham~2002, 
Khos\-ro\-sha\-hi~et~al.~2004 (K04), La~Barbera~et~al.~2005, R05] 
have found that  S\'ersic parameters $\{n, R_{\rm e}, 
\mu_0\}$ of early-type galaxies define a plane in log-space
of the form
\begin{equation}
\log n = a \log R_{\rm e} + b \, \mu_0 + c\,,
\label{eq:php}
\end{equation}
which is termed the ``photometric plane'' (PHP). 
Some authors instead of $\mu_0$ use the mean effective brightness 
 $\langle \mu \rangle_{\rm e}$ (e.g. Graham~2002, La~Barbera~et~al.~2005). 
The different 2D correlations of $\S$\ref{sec:2Dcorrel} can be considered 
then projections of the PHP.

We have computed, by a linear-square fit procedure, the coefficients of
 this PHP for our remnants under the assumption of 
a constant mass-to-light ratio. Since we find that using
 $\langle \mu \rangle_{\rm e}$ leads to about twice the {\sc rms} in $\log n$
than using $\mu_0$, we restrict ourselves to an expression of the 
form (\ref{eq:php}).

%%%%%%%%%%%%%%%%%%%%%%%%%5
\begin{table}
 \centering
 \begin{minipage}{140mm}
  \caption{Photometric Plane coefficients}\label{tab:php}
  \begin{tabular}{lccrc}
  \hline
{\sc id}&  $a$ & $-b$ & $c$ & {\sc rms}$_n$ \\
 \hline
K00      &  $0.17 \pm 0.03$ & $0.069\pm 0.007$ & $1.2\pm 0.1$ & 0.04 \\
D01      &  $0.09 \pm 0.02$ & $0.058\pm 0.004$ & $1.3\pm 0.1$ & 0.06 \\
K04      &  $0.21 \pm 0.09$ & $0.074\pm 0.013$ & $1.7\pm 0.3$ & 0.13 \\
R05E     &  $0.15 \pm 0.02$ & $0.066\pm 0.003$ & $1.1\pm 0.0$ & 0.04 \\
R05dE    &  $0.16 \pm 0.04$ & $0.082\pm 0.004$ & $1.6\pm 0.1$ & 0.07 \\
RJ04     &  $0.11 \pm 0.04$ & $0.054\pm 0.004$ & $1.0\pm 0.0$ & 0.06 \\
M($I_1$) &  $0.05 \pm 0.00$ & $0.057\pm 0.001$ & $-1.0\pm 0.0$ & 0.05 \\
M($I_2$) &  $0.05 \pm 0.01$ & $0.057\pm 0.001$ & $-1.0\pm 0.0$ & 0.05 \\
\hline
\end{tabular}
\end{minipage}
\end{table}
%%%%%%%%%%%%%%%%%%%%%%%%%%%%%%%

%%%%
\begin{figure}
\centering
\includegraphics[width=8.5cm]{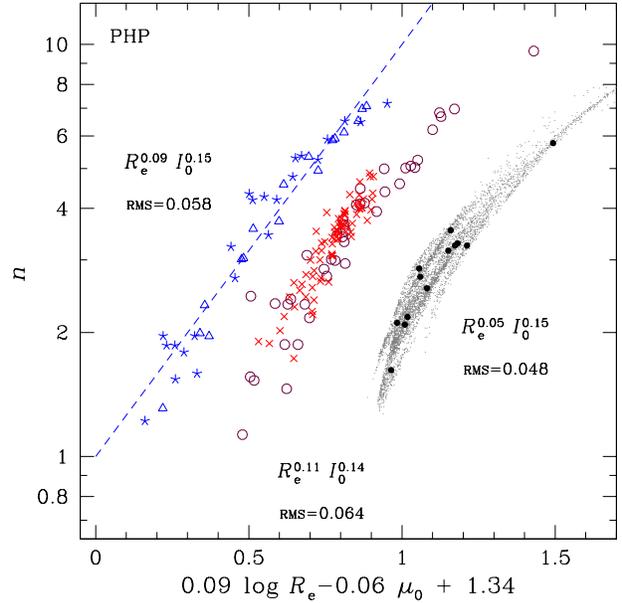}
\vspace{-0.3cm}
\caption{Representation of the photometric plane (PHP) for the same observational  data of Figure~\ref{fig:NMu0correl}, and for our numerical remnants. 
Symbols are as in Figure~\ref{fig:NMu0correl}.  
The $x$-axis is defined here with the values obtained from D01 data.
  Scalings $n \propto R_{\rm e}^\alpha I_0^\beta$ are indicated, along with
the {\sc rms} of the linear fit. $N$-body remnants have been displaced
 a constant term.}
\label{fig:PHPrem}
\end{figure}
%%%%

In Table~\ref{tab:php} we list the values of the coefficients of (\ref{eq:php})
 found in the works of K00, K04, RJ04, the elliptical and dwarf ellipticals 
 of R05, and
those we obtain from the data of D01. Also shown are the coefficients
 found for
our merger remnants for both fitting 
 radial intervals  $I_1$ and $I_2$;
M($I_1$) and M($I_2$), respectively.
In Figure~\ref{fig:PHPrem} we  plot the PHP from these data, using for
illustrative purposes the values from the data of D01 to 
define the abscissa axis.

The remnants' coefficient $b$ in (\ref{eq:php}), associated with $\mu_0$, 
is rather consistent with the observed ones, aside of those found 
in dEs (R05).
The coefficient $a$, associated with $R_{\rm e}$, is less well reproduced.
This is not surprising taking into account 
 the large dispersion 
in the $n$--$R_{\rm e}$ relation (see Figure~\ref{fig:NRcorrel}).
 As several authors have pointed out (e.g. K00, K04, R05)
a slight curvature towards small values of $n$ is observed, a feature that 
tends to be reproduced here by the effect of merger $M11$ that has
 $n\! \approx \! 1.5$.
The {\sc rms} 
 is similar for both the  observational data and our simulations. The
best overall agreement  is obtained with the 
  data of D01.

We consider that our  $a$ and $b$ values are rather
 consistent with the whole set of values
listed in Table~\ref{tab:php}, and that the numerical remnants are able to 
reproduce the PHP. In $\S$\ref{sec:discussion} we argue that the PHP is not
really a plane, but a ``pseudo-plane'' with a small curvature at low
values of $n$ due to the intrinsic properties of  S\'ersic model.

\subsection{Dark Haloes}\label{sec:dark}
%%%%%%%%%%%%%%%%%%%%%%%%%%%%%%%%

Dark haloes in cosmological simulations are started to being described 
by a 3D S\'ersic function 
(Merritt~et~al.~2005, Prada~et~al.~2005, 
Graham~et~al.~2005), of the form
\begin{equation}
\rho(r) = \rho_0 \exp[ -d_n (r/r_{\rm e})^{1/n} ] 
\label{eq:3dsersic}
\end{equation}
with $r$ being the spatial radius, $r_{\rm e}$ a 3D ``effective radius'', and
$d_n \approx 3n-1/3+0.005/n^2$ (Graham~et~al.~2005). 
It has been
found that (\ref{eq:3dsersic}) provides a better fit to dark haloes
than the typical  NFW  or M99 (Moore~et~al.~1999) 
density profiles. 
S\'ersic indices of  about $6$, with a scatter of $\approx 1$, are found for the cosmological dark haloes.

%%%%%%%%%%%%%%%%%%%%%%%%%
\begin{table}
 \centering
 \begin{minipage}{140mm}
  \caption{Dark haloes 3D S\'ersic fits}\label{tab:darkies}
  \begin{tabular}{lccrcc}
  \hline
{\sc id}& $n$ & $r_{\rm e}$ & $\log \rho_0$ &   & {\sc rms} \\
        &     & [kpc]       & [M$_\odot/{\rm kpc}^3$] & &    \\
 \hline
$M01$ & 3.3  & 86.7  &  9.08 &  & 0.05 \\
$M02$ & 3.5  & 28.4  &  9.81 &  & 0.03\\
$M03$ & 4.5  & 26.5  & 10.90 &  & 0.04\\
$M04$ & 4.0  & 42.1  & 10.17 &  & 0.08\\
$M05$ & 4.3  & 20.4  & 10.93 &  & 0.06\\
$M06$ & 4.4  & 25.2  & 10.88 &  & 0.05\\
$M07$ & 3.9  & 23.0  & 10.32 &  & 0.05\\
$M08$ & 3.8  & 28.9  &  9.99 &  & 0.05\\
$M09$ & 4.5  & 29.0  & 10.92 &  & 0.07\\
$M10$ & 4.0  & 31.9  & 10.24 &  & 0.12\\
$M11$ & 3.7  & 41.3  & 10.03 &  & 0.06\\
$M12$ & 3.3  & 39.5  &  9.38 &  & 0.05\\
$MM$  & 3.0  & 82.6  &  8.78 &  & 0.04\\
\hline
\end{tabular}
\end{minipage}
\end{table}
%%%%%%%%%%%%%%%%%%%%%%%%%%%%%%%5

%%%%
\begin{figure}
\centering
\includegraphics[width=8.1cm]{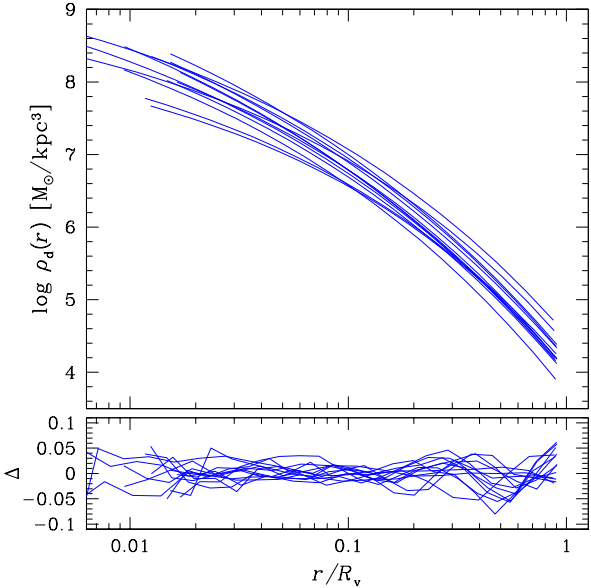}
\vspace{-0.3cm}
\caption{({\it Top}) Fitted 3D S\'ersic profiles (\ref{eq:3dsersic}) to the dark haloes of our  remnants. ({\it Bottom}) Logarithmic residuals of the fits.}
\label{fig:darkprofiles}
\end{figure}
%%%%

We have fitted 3D S\'ersic profiles (\ref{eq:3dsersic})
  to the dark haloes of our remnants. The
radial range of the fit was from the convergence radius $r_{\rm c}$ (e.g. 
 Power~et~al.~2003) to 
the dynamical virial radius of the remnant (Table~\ref{tab:global}). 
However, instead
of using the orbital period at the $r_{200}$ radius
 to determine $r_{\rm c}$ as done in cosmological
simulations,  we used
the orbital period at the virial radius. 

 In Figure~\ref{fig:darkprofiles} 
 we display the fitted 3D S\'ersic profiles, along with their residuals. 
In Table~\ref{tab:darkies}  the values of the 3D S\'ersic 
 parameters are listed for each remnant, as well as the corresponding 
{\sc rms}.
It can be seen that   
the 3D S\'ersic profile  (\ref{eq:3dsersic}) is a very good representation 
of the density distribution up to the virial radius of each remnant. This 
is in  concordance with the behaviour of the 3D S\'ersic profile for 
characterising  cosmological haloes.

The haloes of the $N$-body remnants 
have a mean 3D S\'ersic index $\langle n \rangle\! = \!
3.9 \pm 0.5$; the uncertainty being the standard deviation.
 A value that 
is lower than that found for cosmological haloes. 
However, it is consistent with the mean value of
 the dark haloes of the progenitors $\langle n \rangle\! = \! 3.7\pm
0.3$; as expected from the pre\-ser\-vation of the cuspyness of dark haloes 
in mergers (e.g. Boylan-Kolchin \& Ma~2004, Aceves~\&~Vel\'azquez~2006, 
Kazantzidis, Zentner \& Kravtsov~2006).

%%%%
\begin{figure}
\centering
\includegraphics[width=8cm,height=10cm]{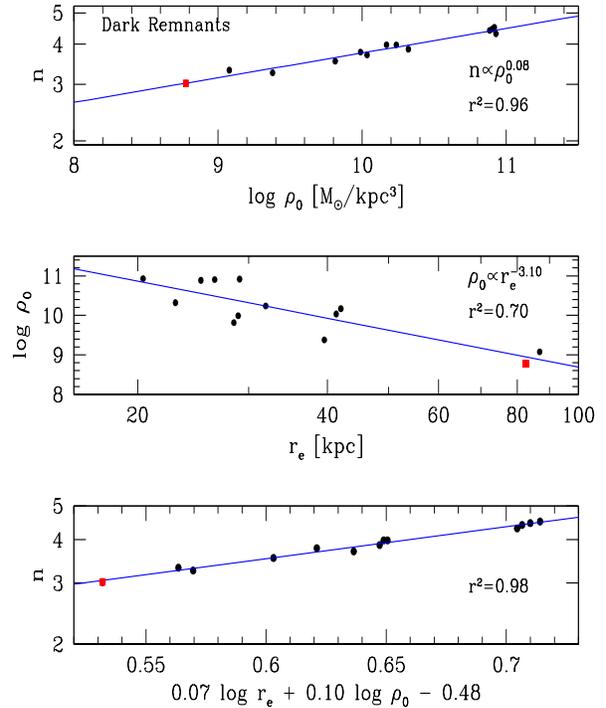}
\vspace{-0.3cm}
\caption{Relations between different physical quantities of the dark matter haloes obtained from a S\'ersic de-projected profile. }
\label{fig:darkies}
\end{figure}
%%%%

Differences in results are expected since  the outer radius of the fits are not the same; we use a dynamical virial radius while for cosmological haloes the fits are done up to  $r_{200}$ or further out (e.g. Prada~et~al.~2005). 
It should be noticed that 
S\'ersic fits (\ref{eq:3dsersic}) by  
Graham~et~al.~(2005) have values for
 $r_{\rm e}$ (see their Table~1) in some cases
larger than their cosmological virial radius; 
these last ones listed in Table~1 of Diemand, 
Moore \& Stadel~(2004).
For example,
 their haloes G02 and G03 have 
$r_{\rm e}\!=\! 391.4$~kpc and 405.6~kpc, respectively, 
while their  virial radii are $337\,$kpc and $299\,$kpc; 
halo B09 shows even a more larger discrepancy.
Unfortunately, they do not provide their numerical
half-mass radii to make a direct comparison with 
the values $r_{\rm e}$ they obtained.
Also, Merritt~et~al.~(2005) and Prada~et~al.~(2005) do not provide the fitted 
 values $r_{\rm e}$ and the numerical half-mass radii. 
This makes uncertain any comparison of our results with these works.

 Figure~\ref{fig:darkies} shows different
 relations among the 3D S\'ersic parameters 
for the haloes of our remnants:
 $n$--$\rho_0$ and $\rho_0$--$r_{\rm e}$, and in analogy to the PHP we have
 constructed a 3D dark S\'ersic plane (DSP). We find, assuming a log-linear
 correlation, 
 that $n\propto \rho_0^{0.08}$ and $\rho_0\propto r_{\rm e}^{-3.10}$ with 
coefficients of determination $0.96$ and $0.70$, respectively.
This indicates that 
 $n$--$\rho_0$ can be considered with confidence a true log-linear positive
correlation, as its was in the corresponding 2D case, but we do not deem
  on the same level $\rho_0$--$r_{\rm e}$. 
 The DSP found for the remnants is
\begin{equation}
\log n \! = \! (0.07 \pm 0.02) \log r_{\rm e} - (0.04 \pm 0.00) \bar{\mu}_0 -
 (0.48 \pm 0.07)\;
\label{eq:dsp}
\end{equation}
where $\bar{\mu}_0\!=\!-2.5 \log \rho_0$. This turns out to be 
a very tight correlation, with a
coefficient of determination of $0.98$ and  {\sc rms} of $0.007$, for the range  of haloes masses considered in our simulations (see Table~\ref{tab:global}).

%%%%%%%%%%%%%%%%%%%%%%%%%%%%%%%%%%%%%
\section{Final Remarks and Conclusions}\label{sec:discussion}
%%%%%%%%%%%%%%%%%%%%%%%%%%%%%%%%%%%%%

The results found here, as well as those of NT and GGB, show that 
the  merger scenario is capable of reproducing 
the S\'ersic properties of observed
 elliptical galaxies. 
This work shows, however, that
 the presence of a ``primordial'' bulge in the progenitors is not 
 necessary to satisfy, for example, the observed values of  
the shape parameter $n$; as was suggested by  NT and GGB.

It is likely that the different results with NT and GGB have their 
 origin on  the initial
properties of the progenitors; in particular,  their
dark matter distribution. We have used cuspy (NFW-type) dark haloes in
contrast to those used by NT (pseudo-isothermal) and GGB (Lowered Evans) that
have  a constant density core. It is probable that the higher concentration of dark matter used here, affected the distribution of luminous matter in a way to increase the index $n$ which is correlated with the luminous concentration (Trujillo~et~al.~2001). Other initial conditions of the progenitors and of the encounters such as energy and angular momentum, both intrinsic and orbital, may have played a role in the final concentration of luminous matter of the remnants, as indicated by the index $n$. A systematic study of the way different dynamical elements determine the S\'ersic index is, however, beyond the scope of the present work.

\begin{figure}
\centering
\includegraphics[width=7cm,height=7cm]{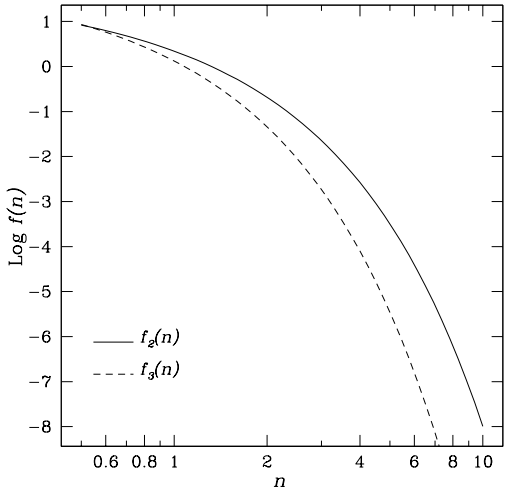}
\vspace{-0.3cm}
\caption{Form factor $f_{2}(n)$ in equation (\ref{eq:plane2}) contributing
to the observed curvature of the photometric ``plane'' at low values of
 $n$; for an {\sl ensemble} of equal total-luminous projected mass galaxies. 
An analogous $f_3(n)$ for a 3D distribution is also plotted. 
 }
\label{fig:Nfactor}
\end{figure}

We have shown that haloes of remnants define a tight dark S\'ersic plane
(DSP) analogous to that of the luminous matter and with less dispersion.
No indication of curvature is present, at difference to what is noted 
 in the PHP at low values of $n$.
 We argue here that this curvature is real and is related to an intrinsic
property of a S\'ersic profile.
Consider the expression for the total luminous
matter associated with a S\'ersic profile (\ref{eq:Lmass}). This
 can be written in log-space as
\begin{equation}
\log L_{\rm T} = \log n - 0.4\, \mu_0 + 2 \log R_{\rm e} + \log f_2(n)
\label{eq:plane2}
\end{equation}
with the ``form factor'' $f_2(n)= 2 \pi \Gamma(2n)/b^{2n}$.
A given set of galaxies with equal $L_{\rm T}$ and different
S\'ersic parameters would define an exact log-plane, except for the presence 
of the $\log f_2(n)$ term. 
In the 3D 
S\'ersic function (\ref{eq:3dsersic}) the analogous form factor is 
 $f_3(n) = 4 \pi \Gamma(3n)/d^{3n}$.
These non-constant terms 
 introduce
a systematic change in a PHP-like expression. 
The importance of the form factor is 
larger for values
 $n\lta 1$ and smaller for $n \gta 1$; 
 both 
 $f_2(n)$ and $f_3(n)$ are shown in 
 Figure~\ref{fig:Nfactor}. Thus, the form factor of the S\'ersic
 model determines the curvature observed in the PHP. 
This explains why no curvature is found by  La~Barbera~et~al.~(2005) whose
galaxies show $n \gta 2$, but this can be seen in dwarf ellipticals with 
 several values of $n\lta 1$ (K04). Also, the DSP does not show such
curvature since $n \gta 3$ (see Figure~\ref{fig:darkies}).
Furthermore, the dispersion
 about these  ``planes'' is determined by the luminosity or dark mass range of 
the galaxy  sample.

%%%%
%%%%

It remains to study the central phase-space densities of the remnants, to see
if they   are consistent with the estimates for ellipticals
(e.g. Carlberg~1986), and to analyse their kinematical properties with those
observed in elliptical galaxies.  We plan to study these topics in a future work. In summary, our main conclusions are as follows:

\begin{enumerate}
\item Collision-less mergers of \emph{pure} disc galaxies yield values and 
  distributions of  S\'ersic parameters consistent with those observed
  for bona fide ellipticals. The existence of a bulge in merging spirals 
 does not appear to be a necessary condition on grounds
  of S\'ersic properties of the remnants.

\item 
 The suggested 
  positive log-linear correlation between 
  the size ($R_{\rm e}$) and structure ($n$)
  in ellipticals is not supported. 
  However, 
  the strong $\log~n$--$\mu_0$ linear correlation found in observational 
  studies is supported by our merger simulations. On other hand, 
  the PHP is fairly well reproduced. For these results 
  a  constant mass-to-light ratio is assumed. 

\item The final dark haloes of remnants show values of $n\approx 4$ lower than
  those found in  cosmological simulations $n\approx 6$. The 
  difference may be attributable to the non-equivalence outer radius, 
  where the dynamical virial radius was used in our case to carry out the 
  fitting by a  S\'ersic profile. 
  Haloes define a tight Dark S\'ersic Plane (DSP) in three
  dimensions, with no indication of curvature at the level of the smaller $n$
  obtained. 

\item The curvature observed in the  PHP 
 at low values of $n$ is an intrinsic manifestation
  of the properties of S\'ersic model, due to the presence 
 of a non-constant term dependent of $n$. 
\end{enumerate}

%%%%%%%%%%%%%%%%
\section*{Acknowledgments}
%%%%%%%%%%%%%%%%%%%%%%%%%%%

This research  was funded by CONACyT-M\'exico project 37506-E. 
We appreciate the kindness of Chazhiyat Ravikumar for providing us 
observational data used in this work.

%%%%%%%%%%%%%%%%
\section*{Appendix}
%%%%%%%%%%%%%%%%%%

We briefly discuss here the effect of the radial interval on the fitting
process of a S\'ersic profile to a mass distribution.
  To do this we generate exact
 S\'ersic profiles with $R_{\rm e}\!=\! 1$, $L_{\rm T}\!=\!1$, and
  $n=2,4,8$. Also, random fractional errors $\le \{1,10,20\}$\% are 
 introduced. 
  The radial fitting interval is chosen as follows.
 A random inner point $\xi$ is selected from the interval
 $[0.03,0.5]R_{\rm e}$ while the outer radius, $\eta$, is randomly generated
from the interval $[R_{70},R_{95}]$; where $R_{70}$ and $R_{95}$ correspond to
 the radii containing 70\% and 95\% percent of the projected mass. These two
points define our radial fitting interval $[\xi,\eta]$.
 To corroborate the importance the importance of the underlying mass
 distribution  we use
 also a Hernquist (1990) mass model with an without errors.

\subsubsection*{S\'ersic Distribution}
%%%%%%%%%%%%%%%%%

Table~\ref{tab:app1} lists the mean 
 values of $\{n,R_{\rm e},\mu_0\}$ obtained from fitting  
 1000 Monte Carlo experiments for three S\'ersic models with errors 
as indicated above. 
Each line lists, in order of the ascending error introduced to the
 theoretical S\'ersic profile, the parameters recovered from the fit inside the random interval 
$[\xi,\eta]$. The standard deviation for each
quantity is provided. 

These results show that the determination of  S\'ersic
parameters is very robust, for errors $\lta 10$\%,  against the size of 
the fitting region. As the error in the ideal S\'ersic  distribution
 increases the dispersion grows. This is more clearly appreciated for 
 index $n$. 
  In the limit of zero error, even for a random radial fitting interval, 
 the model parameters are recovered exactly. For this case, 
 we conclude that the radial fitting range does not
 has an important effect on S\'ersic fitted parameters.

%%%%%%%%%%%%%%%%%%%%%%%%%
\begin{table}
 \centering
 \begin{minipage}{8cm} %{140mm}
  \caption{S\'ersic fits.}\label{tab:app1}
  \begin{tabular}{cccc}
  \hline
 $n_{\rm{true}}$ & $n$ & $R_{\rm e}$ & $-\mu_0$ \\
 \hline
 2 & $2.000\pm 0.015$ & $1.000 \pm 0.003$ & $\hphantom{0}0.956\pm \hphantom{0}0.030$  \\
   & $2.014\pm 0.162$ & $1.002 \pm 0.031$ & $\hphantom{0}0.981 \pm \hphantom{0}0.319$  \\
   & $2.061\pm 0.418$ & $1.009 \pm 0.082$ & $\hphantom{0}1.067 \pm \hphantom{0}0.819$  \\
\hline
 4 & $4.001\pm 0.037$ & $1.000 \pm 0.004$ & $\hphantom{0}4.940 \pm \hphantom{0}0.080$  \\
   & $4.041\pm 0.408$ & $1.001 \pm 0.043$ & $\hphantom{0}5.026 \pm \hphantom{0}0.876$  \\
   & $4.222\pm 1.749$ & $1.010 \pm 0.139$ & $\hphantom{0}5.407 \pm \hphantom{0}3.719$  \\
\hline
 8 & $8.002\pm 0.087$ & $0.999 \pm 0.007$ & $13.261 \pm \hphantom{0}0.196$  \\
   & $8.119\pm 1.042$ & $0.998 \pm 0.071$ & $13.521 \pm \hphantom{0}2.316$  \\
   & $8.663\pm 4.776$ & $1.006 \pm 0.251$ & $14.719 \pm 10.407$  \\
\hline
\end{tabular}
\end{minipage} 
\end{table}
%%%%%%%%%%%%%%%%%%%%%%%%%%%%%%%5

\subsubsection*{Hernquist Distribution}

We now consider that the case where the underlying mass distribution follows
 a Hernquist model. Here, $R_{\rm hl}$ denotes its theoretical projected
 half-light (mass) radius. 
It is found that in the fitting interval 
 $[0.03,2.79]R_{\rm hl}$ , the underlying Hernquist's profile is fitted by 
 a S\'ersic profile with index  $n=2.6$ and $R_{\rm e}=0.82$ in agreement with
 NT. For a radial fitting interval of 
  $[0.03,14.5]R_{\rm hl}$ we find that  
 $n\!=\!3.67$ and $R_{\rm e}\!=\!1.10$. This indicates that the process of
 fitting a S\'ersic profile is far more sensitive when the underlying mass
 distribution does not follows a S\'ersic one.

\begin{figure}
\centering
\includegraphics[width=8.5cm]{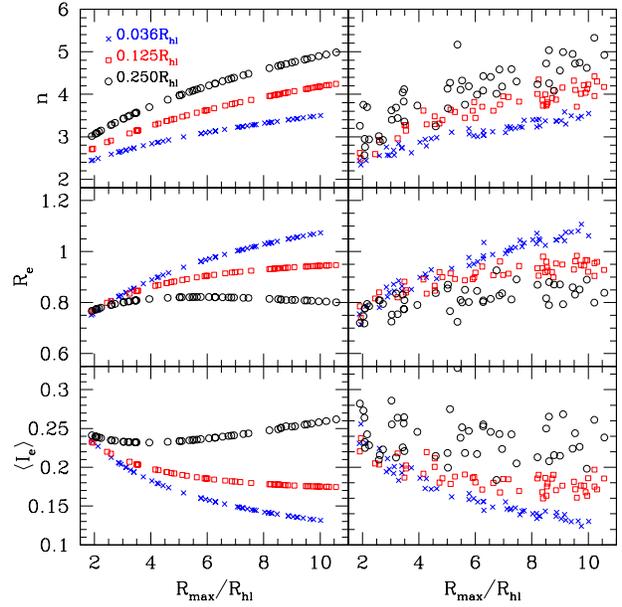}
\vspace{-0.5cm}
\caption{Dependence of S\'ersic parameters $\left\{n,R_{\rm e},
\langle I\rangle_{\rm e}\right\}$ with the fitting interval for
an underlying Hernquist profile.  ({\it Left}) Ideal Hernquist profile. 
({\it Right}) A random
fractional error $\le 10$\% is introduced to Hernquist profile. 
 Each point
corresponds to a different Monte Carlo realisation of the profile. Symbols
represent different inner points to carry out the fittings in terms of the
theoretical half-light radius $R_{\rm hl}$ of Hernquist profile. 
}
\label{fig:app1}
\end{figure}

The above was already noted by  Boylan-Kolchin, Ma \& Quataert~(2005), 
where a systematic change in S\'ersic parameters was found  when trying to
fit a Hernquist profile.
In Figure~\ref{fig:app1} ({\it left}) we reproduce this systematic effect, for
the S\'ersic index $n$, $R_{\rm e}$, and the mean effective
surface brightness $\langle I_{\rm e}\rangle$.
If a random error $\le 10$\% 
is introduced the trend is preserved but a large 
dispersion results; especially as the inner radius of radial interval of the
fit is increased.

\begin{figure}
\centering
\includegraphics[width=8.5cm]{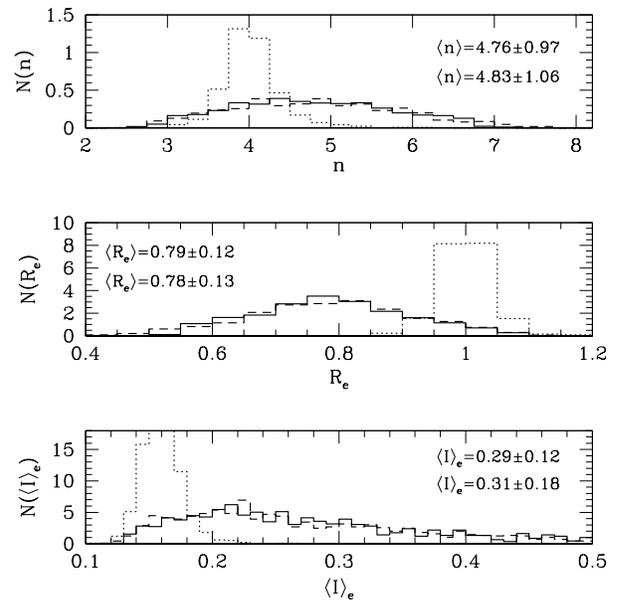}
\vspace{-0.5cm}
\caption{Frequency distributions of fitted S\'ersic parameters
to a Hernquist profile. The radial range of the fit is similar to the interval 
$I_2$ used for our $N$-body remnants. Solid lines is an ideal Hernquist model,
while the dashed line corresponds to the addition of a random error $\le
10$\%.  The dotted line corresponds to a  
 S\'ersic distribution with $n\!=\!4$ and random error $\le 10$\%, shown for comparison. Mean values and standard deviations are indicated. 
}
\label{fig:app2}
\end{figure}

Figure~\ref{fig:app2} shows the distribution of fitted values
 with no error ({\it solid line}) and with a random error 
$\le 10$\% ({\it dashed line}) for an underlying Hernquist profile where 
the radial interval was obtained from
 $\xi \in [0.06,0.50]R_{\rm hl}$ and $\eta\in [\eta_{70},\eta_{95}]$. 
The mean and standard deviations of the distribution are indicated.
For comparison, the histogram of values corresponding to a S\'ersic 
model with $n=4$ with a random error 
$\le 10$\% is also shown ({\it dotted line}); see Table~\ref{tab:app1}, second line in the entry for $n=4$.

%%%%%%%%%%%%%%%
%%%%%%%%%%%%%%

From the above results, it follows that when the underlying mass distribution
 is {\it not} of a S\'ersic type, the fitted values have a rather large
 dispersion even in the presence of no error. 
 In particular, higher values of the index $n$ are obtained for different
 radial ranges of the fit.  
In order to have a confident estimate of $n$, and other parameters,  
one has to sample rather deep inside and outside the luminous (mass)
 distribution; from about $0.1$ to $6\,R_{\rm hl}$.  

In practise, for example, sampling very near the centre of a galaxy may pose
 problems due to resolution effects. This is a particular problem for
 observations of galaxies at different redshifts, using the same angular
 resolution but representing different physical scales, and can lead
 to uncertain S\'ersic parameters.

%%%%%%%%%%%%%%%%%%%%%%%%%%%%%%%%%%%%%%%%%%%%%%%%
%%%%%%%%%%%%%%%%%%%%%%%%%%%%%%%%%%%%%%%%%%%%%%%

%%%%%%%%%%%%%%%%%%%%%%%%%%%

\bsp
\label{lastpage}
%%%%%%%%%%%%%%%%%%%%
\end{document}